# Phase space dynamics and control of the quantum particles associated to hypergraph states

Vesna Berec[1, 2a]

[1]*Institute of Nuclear Sciences, P. O. Box 522, Belgrade, Serbia*
[2]*University of Belgrade, Serbia*

**Abstract.** As today's nanotechnology focus becomes primarily oriented toward production and manipulation of materials at the subatomic level, allowing the performance and complexity of interconnects where the device density accepts more than hundreds devices on a single chip, the manipulation of semiconductor nanostructures at the subatomic level sets its prime tasks on preserving and adequate transmission of information encoded in specified (quantum) states. The presented study employs the quantum communication protocol based on the hypergraph network model where the numerical solutions of equations of motion of quantum particles are associated to vertices (assembled with device chip), which follow specific controllable paths in the phase space. We address these findings towards ultimate quest for prediction and selective control of quantum particle trajectories. In addition, presented protocols could represent valuable tool for reducing background noise and uncertainty in low-dimensional and operationally meaningful, scalable complex systems.

## 1 Introduction

Hypergraph states as a new family of quantum states have recently emerged from quantum information concepts representing an equivalent to equally weighted states simulated via Grover and Deutsch-Joza algorithms [1]. In fact, simple and elegant description of hypergraph states permits their natural implementation in the analysis of quantum algorithms and entanglement purification protocols [2]. Modeling of these states has been used to enhance highly accurate calculations applying a graph theoretic aspect [3, 4] over a broad class of correlated systems. Besides their algorithmic complexity, a specific geometry of hypergraphs, viewed as mathematical objects, can also associate different scales of complexity within elements of a set in the sense of fractal distribution (figure 1).
In this work we demonstrate concept of visualization of quantum dynamics through Wigner function implementing a dynamical phase space description into a set theoretic character of a mathematical hypergraph, which can be further easily encoded into a quantum hypergraph state [2].

---

[a] Corresponding author: bervesn@gmail.com

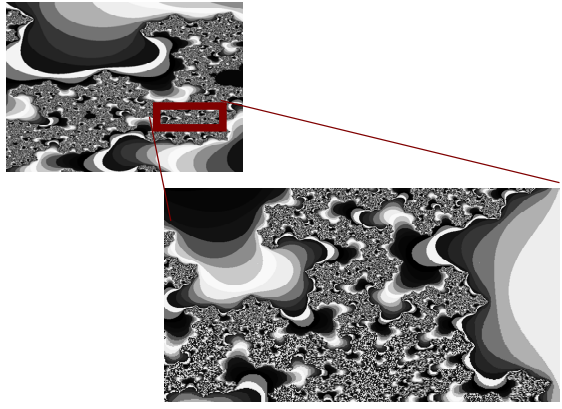

**Figure 1.** Illustration of different scales of complexity and fractal distribution of subsystems within hypergraph set.

## 2 Phase space modelling

In order to present physical property of a quantum particle in the phase space analog to the subspace of Hilbert space, we apply quantum Liouville equation with the zero potential:

$$\frac{\partial P_w(q,p,t)}{\partial t} = -\frac{p}{m}\frac{\partial P_w}{\partial q}, \qquad (1)$$

where $P_w$ denotes particle Wigner distribution in phase space. Former relation is obtained starting from the general description of Wigner function [5, 6, 7], which attains time dependence from the waive function, and assuming Wigner function linear character in the density matrix in case of mixed states:

$$\mathcal{W}(q,p;h) = \frac{1}{\pi\hbar}\int\left\langle q+\frac{\hbar k}{2}\left|\hat{\rho}\right|q-\frac{\hbar k}{2}\right\rangle e^{-ipk/\hbar}dk. \qquad (2)$$

Total volume of the phase space is determined via overall hypergraph boundary $\Gamma$ (which contains maximum collections of graphs). The phase space is discretized into cells of size $\Delta q$ by $\Delta p$, see figure 2., which correspond to a hypergraph vertex set. In this model, each segment of horizontally sliced phase space, with constant momentum over its surface which is at same time equal to a specific hypergraph weight, $W$, evolve independently. Wigner distribution function is associated to each phase space horizontal slice, and obtains a plane wave form: $P_w(q_0,p) = e^{i(kp+\omega t)}$. Higher values of specific hypergraph weights correspond to a higher momentum and higher velocity of the wave. Thus, the specific momentum of each separate horizontal phase space slice relates via specific hypergraph weight with corresponding hypergraph laplacian: $L = 2D_v - \mathbb{H}W\mathbb{H}^T$, where $W$ is the diagonal matrix whose elements are the weights of hyperedges $\omega(e)$; $\mathbb{H}$ denotes incidence matrix [8], and $D_v$ is vertex degree matrix, taken with respect to the total hypergraph boundary ($\Gamma$). According to former momentum dispersion, each wave associated to specific phase space slice, moves at different velocity. In this manner, spectral decomposition of successive discrete frequencies is possible by taking the Fast Fourier Transform (FFT) of Wigner function along individual phase space slices of different momentum $p_i$ at time $t$. Wigner function at time $t+\Delta t$ is obtained using the inverse FFT.

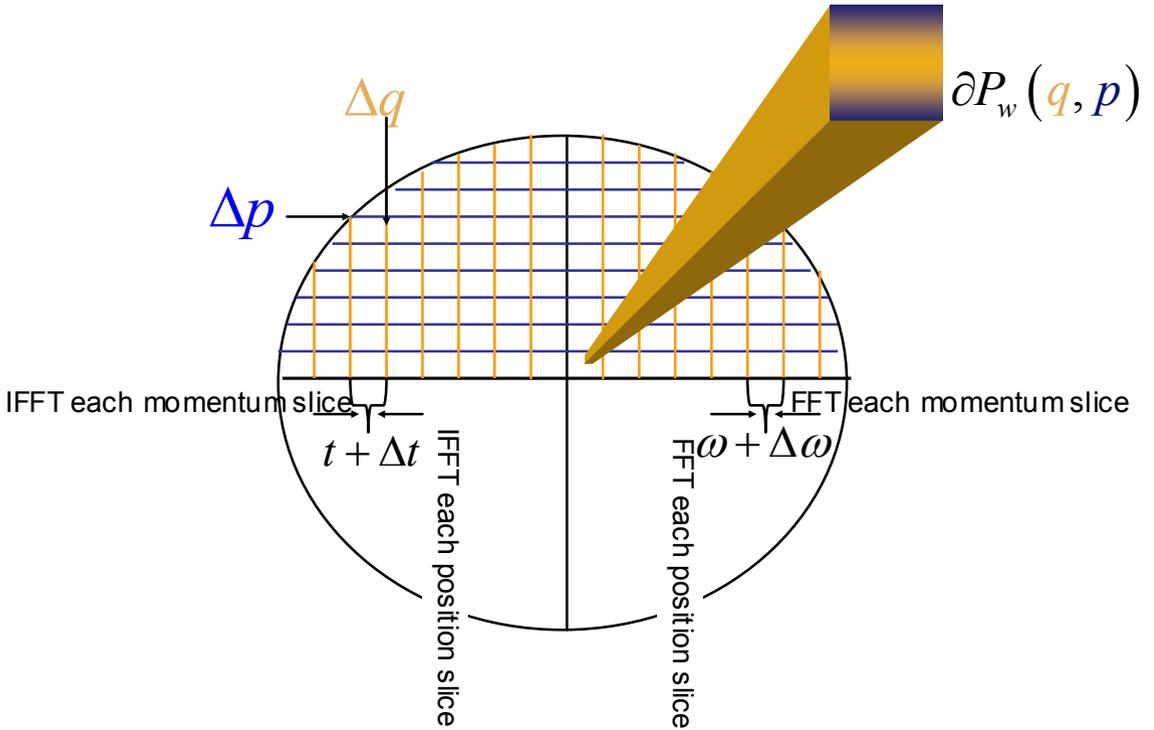

**Figure 2.** Illustration of discretized phase space into cells of size $\Delta q$ by $\Delta p$, with applied spectral decomposition using FFT and IFFT procedures.

Former procedure also holds for the vertical phase space slices. In this case, specific position of each separate slice coincides with specific diagonal elements of hypergraph vertex degree matrix $D_v$ [8] or edge degree matrix $D_e$ (in case if any closed set, belonging to hypergraph, does not contain vertices), which are taken relative to $\Gamma$. In the same manner, particle Wigner distribution is associated to particular phase space vertical slice, i.e., the position slice, taking a plane wave form: $P_w(q, p_0) = e^{i(kq+\omega t)}$. The specific coordinate on each separate vertical slice relates via diagonal weight matrix $f_\omega$ (whose diagonal elements are the sum of the vertices within each hyperedge) with corresponding weighted hypergraph laplacian matrix, in this case of the form: $L = 2D_v - \mathbb{H} f_\omega \mathbb{H}^T$. Likewise in a previous case, as a first step we perform Fast Fourier Transform (FFT) of Wigner function along individual phase space slices: in this case of different specific coordinate $q_j$ at time $t$. The next step is multiplication of a set of resolved frequencies by the each discrete wave number component. At last, the Wigner function in respect to position at time $t + \Delta t$ is obtained by performing inverse FFT over each vertical slice of the phase space.

Following the previous steps in partitioning of the volume of the phase space into horizontal and vertical slices, one obtains the momentum representation and the position representation of the particle Wigner distribution, respectively, simultaneously describing the quantum particle dynamics at a particular instant of time, encodable into hypergraph.

There are several features which hypergraphs adopt that can intrinsically provide an ideal tool for modeling of connectivity information, on a first basis. Namely, a hypergraph is associated to complex structure, see figure 3., which represents a collection of zero or more graphs. Also, hypergraphs vertices may reside simultaneously on different number of graphs. Likewise, hyperedges may originate from multiple vertices or none at all; hyperedges may localize multiple vertices or none at all. Finally, hypergraphs may form complex structures which contain other graphs as vertices.

Besides, a direct transformation to associated hypergraph quantum states is further possible using a class of special quantum Boolean functions, which are independent and commutative [2, 9], i.e., using specific hyperedge gates. We shall further introduce essential elements which are prerequisite for establishing hypergraphs as complex data structures.

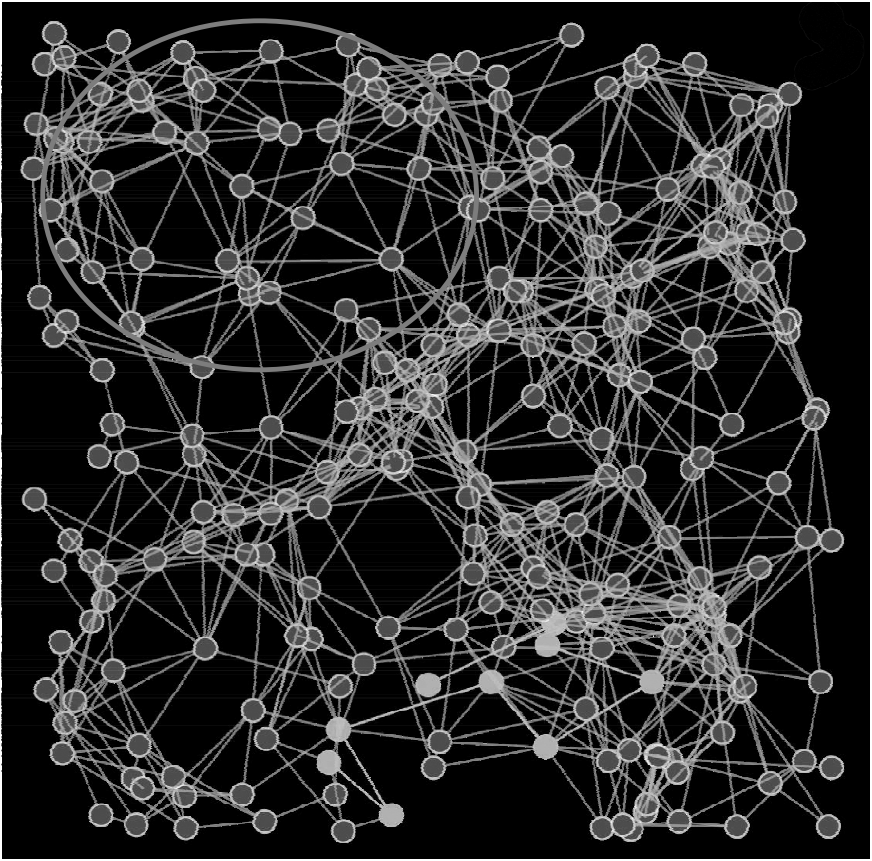

**Figure 3.** A hypergraph based representation of the association between the elements of a vertex set (light grey and dark grey nodes) and their connectivity throughout the net. Individual hypergraph vertices represent specific cells in the circuit and each hyperedge represents a net which forms the circuit's base.

## 2 Hypergraph properties

A weighted hypergraph, $H(V, E, \omega(e))$, see figure 4., represents generalization of the concept of a graph. It is defined by a finite set called the ground set or the vertex set, $V(H)$, whose elements are called vertices, $v$, and by a family of subsets, $E(H)$, whose elements are called hyperedges, $e$.

The degree of a vertex, $v \in V$, represents the number of distinct hyperedges in $E$ that are incident on $v$. For a vertex set $v \in V$ its degree is defined as

$$d(v) = \sum_{\{e \in E | v \in V\}} \omega(e) \cdot h(v, e), \qquad (3)$$

where $\omega(e)$ is a positive number which denotes the weight of the hyperedge $e$, and $h(v,e)$ is the vertex-edge entry element in incidence matrix. For a hyperedge, $e_i \in E$ $(i = 1,...,n)$, its length and degree $d(e_i)$ correspond to its cardinality, $d(e_i) = |e|$. For a hyperedge set, $e \in E$, its degree is given by

$$d(e) = \sum_{e \in E} h(v, e). \qquad (4)$$

The cardinalities of hyperedges can attain different values, from one up to the number of vertices in a hypergraph. In the case when each hyperedge possesses cardinality two, the resulting set system is a graph. $D_v$ is hypergraph diagonal vertex degree matrix whose diagonal element, $d(v_i)$, represents the sum of $i$th row elements of incidence matrix ($\mathbb{H}$), which are multiplied with corresponding weights, $\omega(e_i)$. Correspondingly, $D_e$ is hypergraph diagonal hyperedge degree matrix whose diagonal elements are hyperedges degrees, $d(e_i)$. Vertices $v_i, v_j \in V$ are adjacent in a hypergraph set $H(V,E)$ if, and only if, there exists a hyperedge $e \in E$ such that $v_i, v_j \in e$, i.e., a hyperedge contains these vertices. As a result of this property, a hypergraph $H(V,E)$, $V = \{v_1,...,v_n\}$, $E = \{e_1,...,e_m\}$, can be represented by an $|V| \times |E|$ matrix $\mathbb{H}$, called the incidence matrix, where the elements of $n \times m$ matrix, $\mathbb{H} = (a_{ij})$, are:

$$a_{ij} \begin{cases} 1 & \text{if } v_i \in e_j \\ 0 & \text{otherwise} \end{cases}. \qquad (5)$$

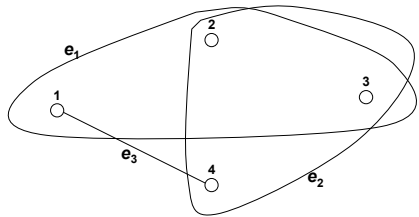

$$\mathbb{H} = \begin{matrix} & e_1 & e_2 & e_3 \\ v_1 & 1 & 0 & 1 \\ v_2 & 1 & 1 & 0 \\ v_3 & 1 & 1 & 0 \\ v_4 & 0 & 1 & 1 \end{matrix}, \qquad D_v = \begin{matrix} & v_1 & v_2 & v_3 & v_4 \\ d(v_1) & 4 & 0 & 0 & 0 \\ d(v_2) & 0 & 3 & 0 & 0 \\ d(v_3) & 0 & 0 & 3 & 0 \\ d(v_4) & 0 & 0 & 0 & 5 \end{matrix},$$

$\omega(e_1) = 1$, $\omega(e_2) = 2$, $\omega(e_3) = 3$, $\omega(e_4) = 4$.

**Figure 4.** Weighted hypergraph: $H = (V, E, \omega(e_i))$, $V = \{v_1, v_2, v_3, v_4\}$, $E = \{\{1, 2, 3\}, \{2, 3, 4\}, \{1, 4\}\} = \{e_1, e_2, e_3\}$ and its incidence matrix $\mathbb{H}$, and vertex degree matrix $D_v$ (where weights, $\omega(e_i)$, correspond to numbers assigned to each hyperedge).

Using above relations, the unnormalized Laplacian, $L$, of a weighted hypergraph is:

$$\begin{aligned} L &= D_v - A \\ &= 2D_v - \mathbb{H}W\mathbb{H}^T, \end{aligned} \tag{6}$$

where $A$ denotes weighted hypergraph adjacency matrix, $A = \mathbb{H}W\mathbb{H}^T - D_v$ [8].

From the aspect of information theory, hypergraphs can be considered as data structures that assign a specific set of related objects (represented by the vertices). The existence of a correlation, interrelated between these objects is represented by a hyperedge. As a result, a scalar weight can be assigned to each hyperedge and each vertex, in order to emphasize the degree of association between the elements of a vertex set that the specific hyperedge contains or connects. The weights on the vertices, for example, present/preserve information about the physical state of specific closed partition (cell), while the weights on the hyperedges contain information about a signal distribution across the circuits (nets).

The part weight function $f_\omega^k = \sum_i f_{\omega_i}$ is defined as the sum of the weights of the vertices, $v \in V$, within the hypergraph part $P_k \in \Pi$ $(k = 1,..., n)$, i.e., it defines the total weight of a given vertex set, $V$. Without loss of generality, the function $f_\omega$ represents the weight of partition ensemble $\Pi = \{P_k\}$, $k = 1,..., n$ of a hypergraph.

## 2.1 Encoding hypergraph into a quantum hypergraph state

Let us start from connected hypergraph, which contains a set of hyperedges, $E(H)$, and a set of vertices, $V(H)$. Note that if $H$ is connected for arbitrary set of $k-1$ hyperedges, we have a $k$-hyperedge connected hypergraph, $H_k(V, E)$ [3, 4]. After assigning a qubit to distinct vertex, each qubit is initialized into a state: $|+\rangle$. A $k$-hyperedge within given set performs controlled-$Z$ (hyperedge gate) operation between specified, or $k$-connected qubits [2, 8], resulting in the following state

$$\begin{aligned} |v_k\rangle &= \prod_{\{i_1,i_2,...,i_k\} \in E} C^k Z_{i_1,i_2,...,i_k} |+\rangle^{\otimes n}, \\ Z_e &= \prod_{\{i_1,i_2,...,i_k\} \in E} C^k Z_{i_1,i_2,...,i_k}, \end{aligned} \tag{7}$$

where the quantum state of $n$ qubits associated with the general $k$-connected hypergraph, $k \in [1, n]$, is denoted as $|v_k\rangle$. When the hyperedge is fixed to $k=1$, this means that $Z_e$ gate induces performance over individual single qubits; the case when the hyperedge is fixed to $k=2$ is analog to a $Z_e$ gate operation over the graph states, while for the case $k = n$, $Z_e$ gate(s) induce performance over all $n$ qubits, corresponding to a hypergraph state [9] normalized via overall hypergraph boundary $\Gamma$ (as shown in figure 5.)

$$\begin{aligned} \Gamma |+\rangle^{\otimes n} &= \prod_{e_1 \in E}^{e_n} Z_e |+\rangle^{\otimes n} = |\varphi_{f_\omega}\rangle, \\ \Gamma\left(\Pi_{\{P_k\}}\right)|+\rangle^{\otimes n} &= |\varphi_{f_\omega}\rangle_k, \quad k = 1,...,n. \end{aligned} \tag{8}$$

where $|\varphi_{f_\omega}\rangle_k$ are real equally weighted states addressed to corresponding sets from hypergraph partition ensemble, $\Pi = \{P_k\}$, $k = 1,...,n$, where $\Gamma(\Pi_{\{P_k\}}) = \sum_{e_i \in E} C^k(d(e_i)-1)$ assigns each hyperedge, $e_i$, which contributes via $C^k(d(e_i)-1)$ to the cost, $\Gamma(\Pi)$, of the hypergraph partition, $\Pi$.

Considering that $P(V)$ is the connectivity power of the set of $V$, which also corresponds to cardinality of a hypergraph, the weight function for a vertex subset $(V' \subset V)$ $f_\omega^k : P(V) \to Z_e$ is such that for $v \subseteq V$, the function $f_{\omega_i}(v)$ defines the weight of a given subset $V'$ of a vertex set, $V$. The function $f_\omega^k(v)$ is then used to define the weight of each hypergraph part $P \in \Pi$, where $\Pi$ is a hypergraph partition $\Pi \subset P(V)$ of the vertex set $V$ (to which individual qubits are assigned) such that $P \cap P' = \emptyset$ and $P, P' \in \Pi$ $(P \neq P')$, addressing to each hypergraph part the quantum Boolean functions as: $|f_\omega\rangle_k = 2^{-\frac{n}{2}} \sum_{v=0}^{2^n-1} (-1)^{f_{\omega_i}(v)} |v\rangle$, where $|v\rangle$ denotes operational basis states for qubits. In order to create real equally weighed states, applicable in Grover and Deutsch-Joza algorithms, a balance between the weights of different parts within the partition of hypergraph state must be established, i.e., partitioning should ensure parts $\Pi = \{P_k\}$, $k = 1,...,n$ with assigned individual part weights $f_\omega^k = f(P_k)$, $1 \leq k \leq n$ such that: $f_\omega^k < (1+\delta)\overline{f}$ holds for qubits $1 \leq k \leq n$, where the average part weight $\overline{f}$ of a partition $\Pi \subset P(V)$ is given by $\overline{f} = \sum_{i=1}^n f_\omega^k / n$, and $f_\omega^k$ weight does not exceed $\overline{f}$ weight by more than a factor of $\delta$, $0 < \delta < 1$.

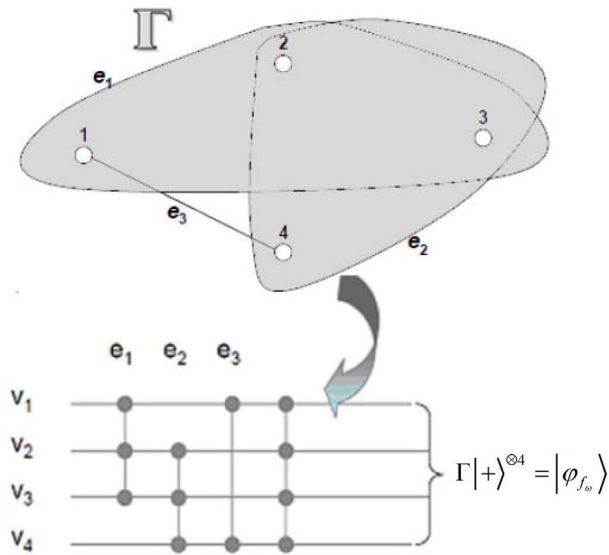

**Figure 5.** Schema for encoding hypergraph into to a hypergraph state obtained applying the set of transformations: $C^1 Z_{123}$, $C^2 Z_{234}$, $C^3 Z_{14}$, $C^4_{1234}$ over $\Gamma$.

## 3 Summary

We have described categorical relation of hypergraphs associated to inherently quantum representation of phase space, where internal components that constitute the building blocks of the dynamical space are defined in terms of their set theoretic (graph) properties. As a final result, in every instant of time we have obtained an intersection of two spectral distributions: velocity distribution in phase space (distribution over momentum) and position distribution in phase space, which correspond to variables $W$ and $f_\omega$, respectively, directly relating to mathematical hypergraph that can be encoded into a quantum hypergraph state.

## References

1. D. Collins, K. W. Kim, and W. C. Holton, Phys. Rev. A **58**, R1633(R) (1998).
2. O. Gühne, M. Cuquet, F.E.S. Steinhoff, T. Moroder, M. Rossi, D. Bruß, B. Kraus and C. Macchiavello, J. Phys. A: Math. Theor. **47**, 335303 (2014).
3. C. Berge, *Hypergraphs, Combinatorics of Finite Sets* (North-Holland Mathematical Library, Volume 45, 1989).
4. A. Frank, *Connections in combinatorial optimization* (Oxford University Press, Oxford, 2011).
5. F. J. Narcowich and R. F. O'Connell, Phys. Rev. A **34**(1) (1986).
6. K. Singer, Molecular Physics, **85**(4), 701-709 (1995).
7. D. T Smithey, M. Beck, M. G. Raymer and A. Faridani, Phys. Rev. Lett., **70**(9), 1244-124 (1993).
8. X. Jiang and N. Petkov (Eds.), *Computer Analysis of Images and Patterns* (Springer-Verlag Berlin Heidelberg, 2009).
9. R.Q. J. Wang, Z. Li, and Y. Bao, Phys. Rev. A **87**, 022311 (2013).